# Determination of centrality in nucleus-nucleus collisions on the MPD/NICA installation

A.Kurepin[1] *, A.Litvinenko[2], E.Litvinenko[2]

[1]Institute for Nuclear Research RAS, Moscow

[2]Joint Institute for Nuclear Research, Dubna

**Abstract.** Measurement of the number of spectators in nucleus-nucleus collisions could be used to determine the number of participant nucleons involved in the interaction, i.e. get information about collision centrality. However, at energies of the NICA collider, energy resolution of the forward hadron calorimeter is insufficient for separation amplitude corresponding to different number of spectators. Uncertainty in the number of spectators leads to a large error in determining centrality. For central events inaccuracy in determining the collision parameter is about 40%, for peripheral ones, no better than 30% at a beam energy of 2.5 AGeV. The multiplicity scintillator detector will allow to obtain more accurate data. A method is proposed for determining the centrality, taking into account the real MPD installation geometry.

## 1 Introduction

The concept of centrality in the collision of relativistic heavy nuclei is an important image that plays the most fundamental role in the collision of nuclear matter with a high baryon density, i.e. at the energies of the experimental FAIR and NICA complexes. In this case, the role of multiple rescattering and secondary processes with the production of pions and kaons is much smaller, and it can be expected that the impact parameter more clearly defines the centrality of the collision, i.e. the role of fluctuations due to secondary interactions is much smaller. The observed particle yields due to collisions of nuclei can be more definitely associated with the magnitude of the impact parameter, i.e. with the centrality of the collision.

It is obvious that the expected formation of a new type of nuclear matter, quark gluon plasma, is more likely in more central collisions. Therefore, a more precise definition of centrality will allow the separation of the formation of a quark gluon plasma and secondary peripheral interactions of nuclei. On the other hand, a higher baryon density in the region of overlapping collisions of nuclei imposes a restriction on the multiplicity of particles produced, which can significantly reduce the statistical accuracy of the determination of centrality and the impact parameter.

The most common method for determining centrality, which is used both when working on a fixed target and at colliders, is to measure the multiplicity of charged particles produced in a collision. The resulting distribution can be divided into several intervals by the magnitude of the multiplicity in each event from central to peripheral. The form of the obtained dependence, as shown in Figure 1, for the collision of xenon nuclei on the ALICE facility and the AD detector, is well described using the Glauber model. Similar multiplicity distributions were obtained for each run of lead collisions using the V0 scintillator detector.

_______________________________________________

∗e-mail: kurepin@inr.ru

As another method for determining centrality, it was proposed to register non-interacting nucleon-spectators using a hadron calorimeter. In this case, the number of interacting nucleons is obtained as the difference between the number of nucleons in the incident nucleus and the number of spectators. This method is most widely used when working with fixed targets, for example, in an experiment NA61.

To simulate such type of forward hadron calorimeters, it is necessary to use programs that take into account the formation of spectators in the collision of nuclei and their passage through nuclear matter. From such programs, one can point to LAQGSM [1] used in this work and SHIELD [2].

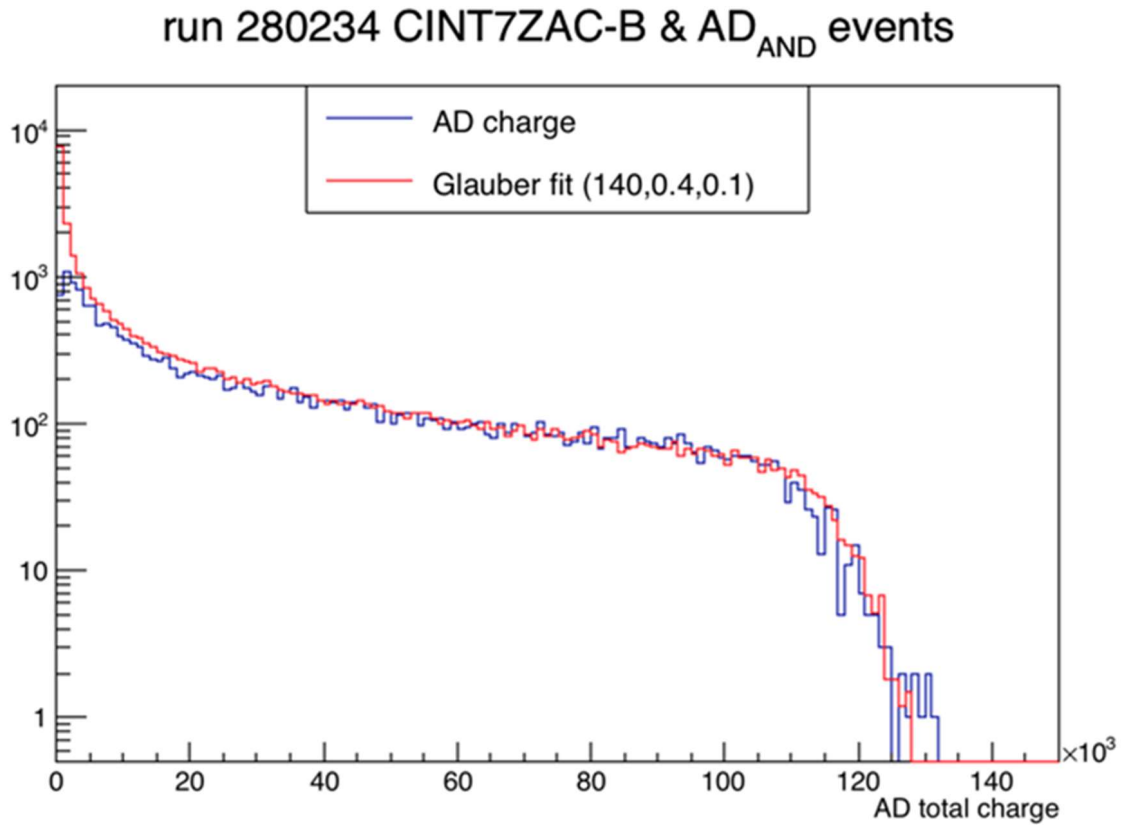

**Figure 1.** Multiplicity distribution for Xe – Xe collisions at 5 TeV

## 2 Centrality determination at MPD/NICA

The use of the forward hadron calorimeter at intermediate energy colliders, as was first shown in Ref. [3], presents difficulties associated with the ambiguity of the information obtained. The presence of a hole in the calorimeter for the passage of the beam leads to the fact that part of the spectators goes into the vacuum tube of the collider. Figure 2 shows the dependence of the spectator yield on the impact parameter. It can be seen that ambiguity is observed in the dependence of the yield on the impact parameter, which does not allow determining the degree of centrality by the magnitude of the measured amplitude of the spectators. In the article [4], a method is proposed for eliminating ambiguity by determining the asymmetry of the yield of spectators according to the data of the internal and external part of the calorimeter.

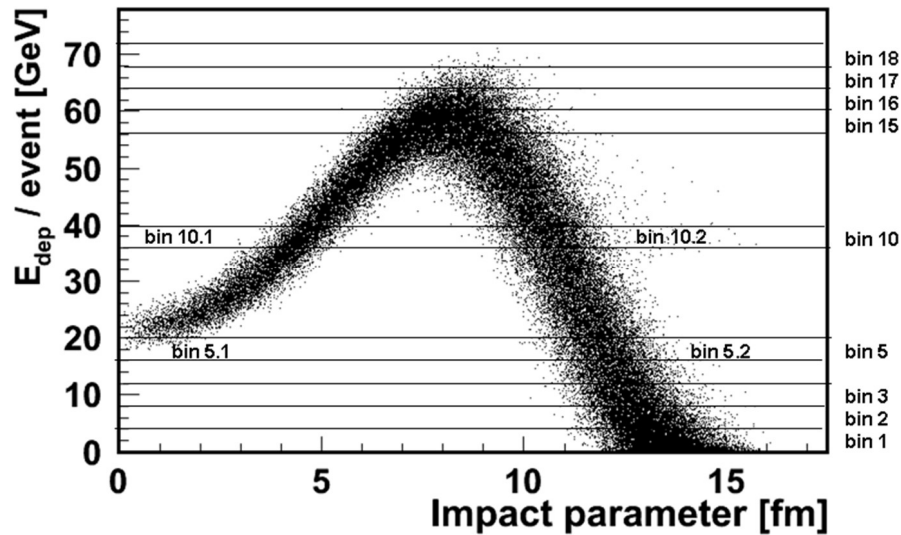

**Figure 2.** Dependence of total energy of spectator deposited in calorimeter on the impact parameter

Another problem when using the forward hadron calorimeter to determine centrality at energies up to 20 GeV is the lack of accuracy in determining the impact parameter. Indeed, to measure the number of interacting nucleons, as mentioned earlier, it is necessary to determine the number of spectators in each collision. Insufficient energy resolution of the calorimeter, obtained from calibration results for single muons and protons, does not allow separation of the amplitudes, corresponding to the individual particles, i.e. determine their number in each collision. This number can be obtained from the calorimeter total absorbed energy value.

In that case, when the resolution allows to separate individual spectators, i.e. determine their number in each event, the relative error of the number of spectators is proportional to $1/\sqrt{N}$, and the same dependence of the error is in determining the centrality or impact parameter b.

But the impossibility of improving the resolution of the calorimeter does not allow to observe maxima in the amplitude of the calorimeter signal corresponding to a different number of spectators in the event, even at NA61 (20 - 70 GeV). In this case, under conditions of insufficient resolution, particularly at the lower energies of NICA ( 2.5 – 5.5 GeV ), the number of spectators can be found from the value of the amplitude of the calorimeter signal, which corresponds to the total energy E absorbed in the calorimeter from some average number of N spectators at a certain value the impact parameter b and at an average value of $E_s$ the energy released in the calorimeter per spectator:

$$E = \sum E_i \approx N \times E_s \qquad (1)$$

According to the calibration results for single muons and protons, the resolution is:

$$\Delta E_s / E_s = 0.56 / \sqrt{E_s} \qquad (2)$$

For the number of spectators in the event it is accepted:

$$N_s = E / E_{s0}$$

where $E_{s0}$ is the energy of a single beam nucleon absorbed in the calorimeter.

To determine the accuracy of $N_s$ and the corresponding parameter b, one must find an error in determining the total energy E.

The error of product is:

$$\Delta E = \partial E / \partial N \times \Delta N + \partial E / \partial E_s \times \Delta E_s = E_s \times \Delta N + N \times \Delta E_s$$

Neglecting the products of various random errors, since
$$lim\ (\Delta N \times \Delta E_s) \to 0$$
we obtain the square of the mean-square error of the total energy:
$$(\Delta E)^2 = E_s^2 \times \Delta N^2 + N^2 \times \Delta E_s^2$$
We divide by:
$$E^2 \approx (N \times E_s)^2$$
Relative error in determining the total energy is:
$$(\Delta E / E)^2 = (\Delta N / N)^2 + (\Delta E_s / E_s)^2$$
According to Poisson:
$$\Delta N / N = 1/\sqrt{N}$$

Taking into account the resolution for a single spectator (2), we obtain for the relative error in determining the number of spectators $N_s = E / E_{s0}$

$$\Delta N_s / N_s = \Delta E / E = \sqrt{(1/N + (0.56/\sqrt{E_s})^2)}$$

For $\sqrt{s} = 5$ GeV $E_{s0} = 2.5$ GeV, $\Delta E_s / E_s = 0.35$
For N = 200 ( peripheral Pb ) $\Delta N_s / N_s = 0.36$
At N = 10 ( central ) $\Delta N_s / N_s = 0.47$

It is wrong to fix the expected number of spectators [4]. Then for N = 10 one get:
$$\Delta N_s / N_s = \Delta E / E = (1/\sqrt{N}) \times (0.56/\sqrt{E_s}) \approx 0.11 = 11\%$$

The difference with the correct estimate is more than 4 times.

Thus, FHCal at MPD/NICA gives the determination of centrality with an accuracy of about 47% for central and 36 % for peripheral events at an energy of $\sqrt{s} = 5$ GeV. At the same time, an additional error for peripheral events is introduced due to the loss of a part of the spectators at mall emission angles.

So determination of centrality with the necessary accuracy in the whole interval of MPD/NICA energies with the help of the calorimeter FHCal is impossible. For the upgrade the FMD (Forward Multiplicity Detector) - a sector scintillator detector of multiplicity - should be installed, which allows to determine the impact parameter b with an accuracy better than 10% in the entire range of centrality.

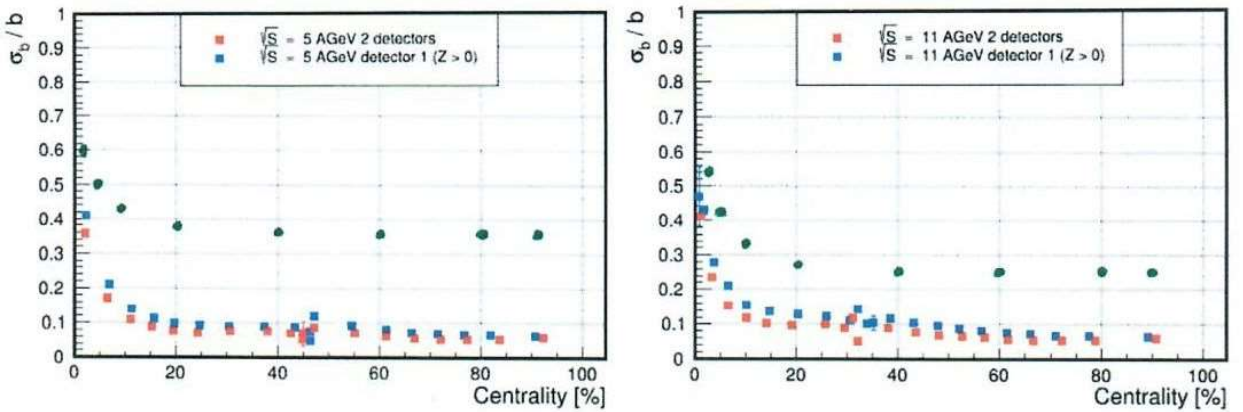

**Figure 3.** Resolution of impact parameter obtained in separate bins of the energy depositions in FHCal for beam energy 2.5 GeV (left) and 5.5 GeV (right). Blue and red points correspond to the estimate with fixed number of spectators. Green points take into account the fluctuation of spectators number

Figure 3 shows the result of calculating the relative error in determining the collision parameter depending on the centrality of the interaction, which is divided into 5 intervals corresponding to the magnitude of the signal amplitude in the hadron calorimeter.

As shown in Ref. [5], the method of determining centrality from the multiplicity of events cannot also be directly used at the energies NICA. Due to the relatively large transverse momentum of the spectators compared to the longitudinal pulse, some spectators enter the multiplicity detector, which leads to the superposition of the number of the most central and most peripheral events. Figure 4 shows the results of calculating the distribution of the multiplicity and total energy by the program LAQGSM separately for protons, which are mainly spectators, pions and fragments with a charge of more than two for the scintillator detector with outer radius 70 cm and inner 6 cm. It is seen that only for pions there is no distortion of the spectrum, and this distribution can be used to determine centrality, provided that the spectrum of charged particles is cleared from the spectators.

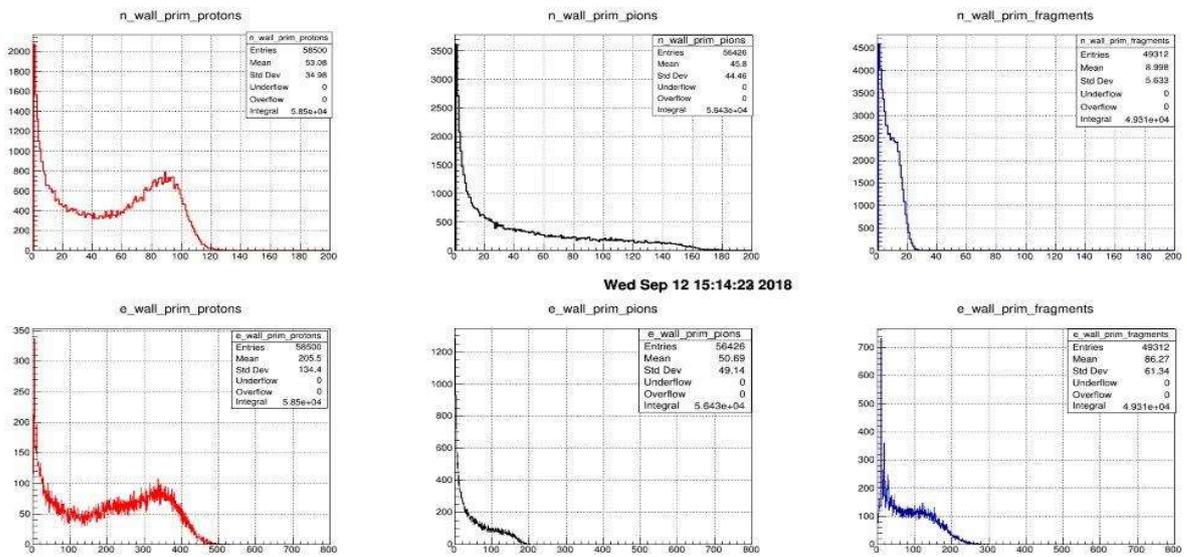

**Figure 4.** Multiplicity distributions for protons, pions and fragments for 5.5 A GeV beam of Au ( upper part ) and total energy distribution in GeV ( lower part )

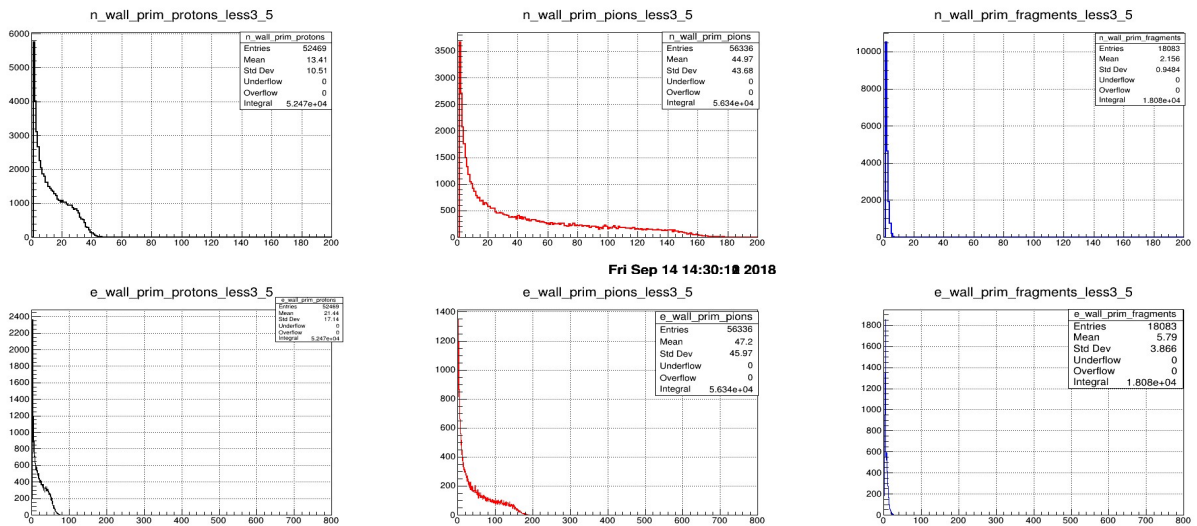

**Figure 5.** Same multiplicity and energy distributions for protons, pions and fragments as in Fig. 4, with condition that all charged particle with energy larger than 3.5 GeV are removed by anticoincidence with hadron calorimeter.

A method is proposed for eliminating spectators by incorporating an anticoincidence of the multiplicity detector with a hadron calorimeter. Pions mostly have energies less than 1.5 GeV. Therefore, setting the anti-coincidence threshold at 3.5 GeV (see Figure 5) almost completely removes spectators, practically without changing the pion spectrum. At the same time, there is no need to achieve high resolution of the calorimeter, and the length of its modules can be reduced. Accordingly, the weight of the calorimeter on each side of 9 tons, which creates additional difficulties for the assembly and disassembly of the installation can be significantly reduced. The accuracy of determining the collision parameter at various centralities is determined by the multiplicity mainly of pions, and is no worse than 10 percent even at an energy of 5 GeV.

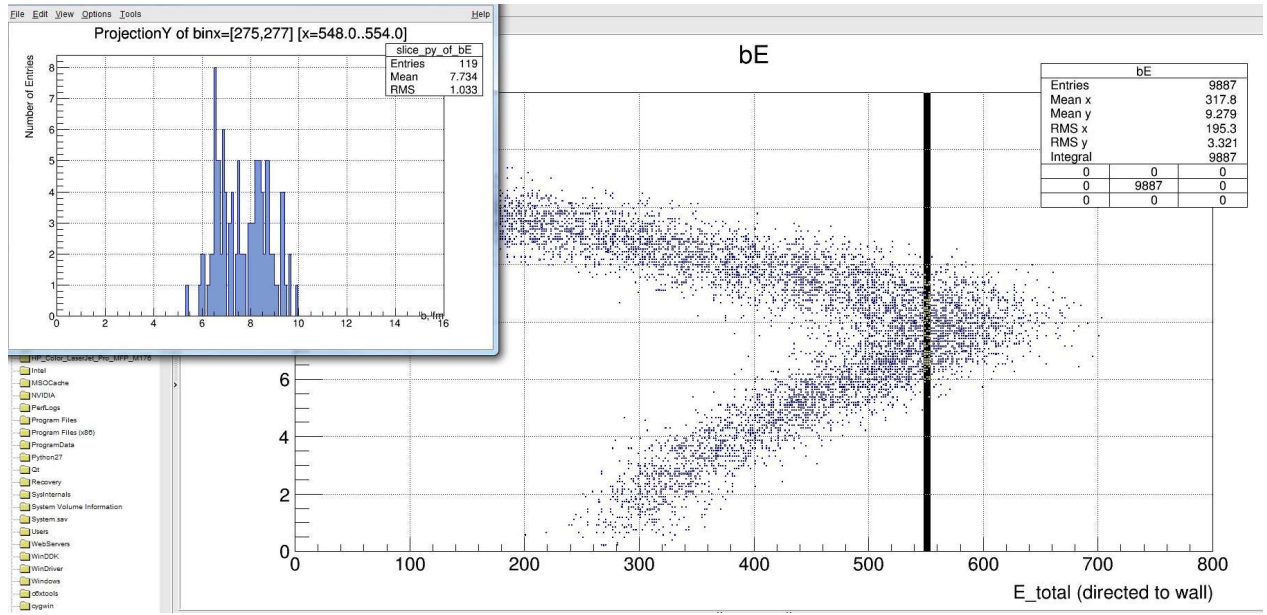

**Figure 6**. Impact parameter distribution for fixed value 550 MeV absorbed in the calorimeter

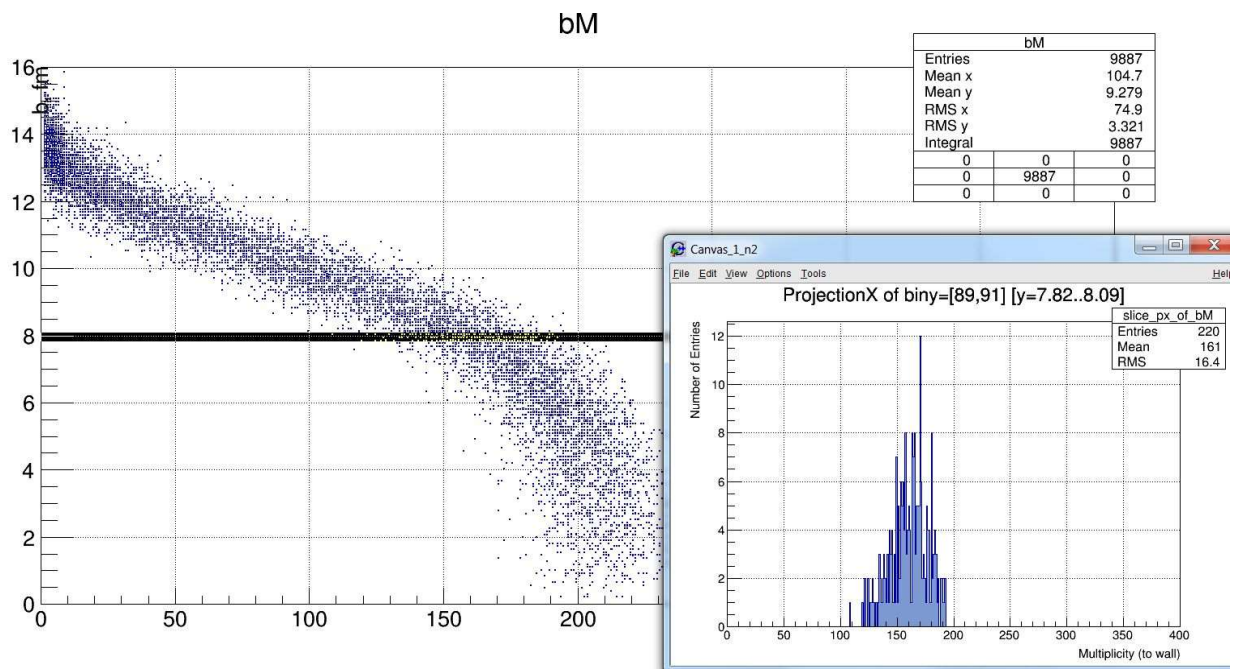

**Figure 7**. Multiplicity distribution for fixed value 8 fm of the impact parameter

As mentioned earlier, due to secondary collisions during the passage of produced particles through nuclear matter, a fatal inaccuracy arises in determining the impact parameter. Figure 6 shows the spread of values of the impact parameter at a fixed value of the energy of the spectators, and

Figure 7 at the multiplicity of the event with the fixed value of the impact parameter. It can be seen that the error in both cases is about 10 percent.

## 3 Conclusion

In conclusion, it can be argued that in order to reliably and accurately determine the centrality and the impact parameter, the MPD installation should also have a scintillator multiplicity detector installed in front of the hadron calorimeter.